\documentclass[prc,twocolumn,showpacs,preprintnumbers,amsmath,amssymb]{revtex4}
\usepackage{graphicx}
\usepackage{dcolumn}
\usepackage{bm}
\input{epsf.tex}

\def\be{\begin{equation}}
\def\ee{\end{equation}}
\def\ba{\begin{array}}
\def\ea{\end{array}}
\def\bea{\begin{eqnarray}}
\def\eea{\end{eqnarray}}
\begin{document}
\title{Reaction and structure effects of light mass nuclei using Glauber model with relativistic and
non relativistic effective interaction densities}
\author{Mahesh K. Sharma$^{1}$} \author{R. N. Panda$^{2}$} \author{Manoj K. Sharma$^{1}$} \author{S. K. Patra$^{3}$}
\affiliation{$^{1}$School of Physics and Materials Science, Thapar
University, Patiala - 147 004, India.}
\affiliation{$^{2}$Department of Physics, ITER, Siksha O Anusandhan University, Bhubneswar-751 030, India}
\affiliation{$^{3}$Institute of Physics, Sachivalaya marg Bhubneshwar-751 005, India}
\date{\today}

\begin{abstract}
We study the structural properties of some nuclei in the region of light mass using simple effective interaction in the frame work of
microscopic non relativistic Hartree-Fock and relativistic mean field formalisms. For the reaction dynamics,
well known Glauber model is used with the conjunctions of the densities obtained from these formalisms.
We observe good agreement of results by both the formalisms in comparison to the experimental values. These two different
approaches seems to be equally capable to reproduce the ground state properties of nuclei in this region. A careful study
of reaction dynamics suggest the superiority of relativistic mean field over the Hartree-Fock with simple effective
interaction densities. But good agreement of reaction
cross section is appeared with simple effective interaction density for halo nuclear systems. The possibility of existence of bubble
nuclei are also studied using both the formalisms.

\end{abstract}

\maketitle

\section{INTRODUCTION}
The total reaction/interaction cross section measurements of nuclei at intermediate energy range has been
extensively used to investigate the size of nucleus and distribution of nucleon in side the nucleus
by the nuclear science community from more than one century \cite{brick1938,bethe1940,nix1945,watson1952,glauber1955,swann1955,wong1974,eck1980, khan1985,tani92, abu2013,shukla2007,patra2009,horiuchi2010,minimo2012, wimmer2012}.
The availability of heavy ion
beams in intermediate energy range makes possible to explore the structures of nuclei over the
wide range in nuclear chart. The advancement in radioactive
ion beam facility (RIB) in this energy range at various laboratories over
the globe open up new opportunities to investigate some new exotic phenomenon.
The study of such exotic nuclei specially those towards the dripline and
the investigation of extended dripline are some of the
current interest \cite{patra2010,saxena2013,singh13}. Although earlier the dripline is well established for $^{24}O$ which is a doubly magic nuclei.
The neutron separation energy and interaction measurement for $p-sd$ and $sd$ shell region suggest the new magic number of
N=16 \cite{ozawa2000} which is supported by Hoffman in Ref. \cite{hoff2008}. The measurement of $^{40}$Mg and $^{42}$Al \cite{baumann07}
beyond the dripline given by various mass formulae challenge the earlier predictions and still the investigation of dripline
is a new challenge for both the nuclear theoretician and experimentalist.
Apart from it the one most exotic phenomenon is the investigation of neutron and proton halo
feature of some nuclei, which occurs due to the extremely weak bound neutrons that decouple from the
nuclear core \cite{tani92,ozawa2001}.
Recently, a new member $^{31}Ne$ is included in the family of neutron
halo \cite{Hama2010,takechi2012}. The isotope $^{31}Ne$ having N=21, breaks
the shell closure structure and in consequence, a large amount of deformation
gets associated with the strong intruder configuration and eventually this nuclear
system lie at island of inversion (IOI) \cite{war1990}. The investigation of nuclear systems at IOI is also be one
of the current interest.
The measurement of nuclear reaction cross-section for
$^{19,20,22}C$ \cite{tana10,Koba2011} shows that the drip-line nucleus
$^{22}C$ has a halo. The one and two neutron removal cross sections
and momentum distribution also support the halo behavior of $^{22}C$.
The $^{22}C$ has N = 16 which is a new magic number in neutron-rich
nuclei \cite{ozawa2000,Tani2001} and forms a Borromean halo structure
($^{21}C$ is unstable).The direct time of flight mass measurement technique suggest $^{19}$B, $^{22}$C,
$^{29}$F and $^{34}$Na are as Borromean dripline nuclei \cite{gaud2012}.\\
Another interesting phenomenon has been found in these days is the bubble structure of some nuclei
like $^{22}$O, $^{23}$F, $^{34}$Si, $^{36}$S, $^{36}$Ar, $^{46}$Ar, $^{84}Se$, $^{134}Ce$, $^{174}$Yb, $^{200}$Hg etc. \cite{grasso2009, Davies1972}.
The idea of bubble effect had been given by Wilson \cite{wilson46}. The possibility of formation of
bubble nuclei has been studied by different nuclear models.
The microscopic calculations using
Skyrme Hartree-Fock (SHF) formalism have been carried for the investigation of this effect in Refs. \cite{campi1973,bender03,decharge03}.
Recently the relativistic and non relativistic mean field formalisms are used to investigate bubble effect in
light \cite{shukla2011} and superheavy region \cite{singh2013}. This effect also influence the reaction cross section of exotic nuclear systems.\\
The work done in Ref. \cite{mahe2013} using the recently proposed simple effective interaction (SEI) in non relativistic
mean field approach\cite{bhuyan12}, successfully shows the good results of reaction cross sections for halo
nuclear systems, which motivate us to see the reaction cross section for low mass region using this approach.
For comparison, we used the well known relativistic mean field (RMF) densities using NL3 parameter set.\\
The paper is organized as, Section I consists of a brief introduction and Section II contains
description of formalisms for relativistic and non relativistic mean field along with the
Glauber model. The calculations and results are presented in Section III. The predictability of HF(SEI) model to explore the
different features of nuclei are included in this section. Finally the summary and conclusions are outlined in Section IV.

\section{The Formalism}
The bulk properties are very important for the study of the characteristics of nuclei. In
this regards, well known relativistic mean field (RMF)
and non relativistic mean field with newly developed simple effective interaction (SEI) are used.
The well known Glauber approach is used here for the study of reaction dynamics for some light mass
nuclei \cite{glau55,Karo75,Chauvin1983}.
The study of reaction
dynamics by this approach strongly depends on the densities of the projectile
and target nuclei. We use densities from axially deformed (Def.) as well as spherically (Sph.) symmetric
RMF formalisms. As we seen in our earlier calculation \cite{mah2013}, the reaction cross section obtained from
densities are parameters dependent upto some extent.
So we used here, same set of parameter NL3 \cite{lala97} in order to see the effect of
deformations on reaction dynamics. The deformation effect is included in $\sigma_R$ through
densities of projectile and target nuclei by taking axially deformed RMF(NL3) densities. The same reaction
study has been done by taking spherically symmetric densities from similar formalism for the sake of comparisons.
We compare the results obtained from both relativistic mean field and non relativistic HF(SEI) densities.
The simple effective interaction (SEI) with parametrization \cite{bhuyan12} are developed with non relativistic microscopic Hartree-Fock
mean field theory. The detail about these formalisms and interaction parameters can be found in
Refs. \cite{lala97} and \cite{bhuyan12} for RMF and SEI, respectively.
Here we used the densities from two different microscopic mean field formalisms to study their effect on the reaction dynamics.

\subsection{Hartree-Fock with simple effective interactions}
The simple effective interaction (SEI) alike to a hybrid of Gogny and Skyrme is used to study the bulk properties of
finite nuclei with in the frame work of Hartree-Fock (HF) formalism. The form of simple effective
interaction (SEI) is given by \cite{bhuyan12}
\begin{eqnarray}
\nu_{eff}(r)&=& t_0(1+x_0P_{\sigma})\delta(r)\nonumber\\
&&+t_3(1+x_3P_{\sigma})({\frac{\rho(R)}{1+b\rho(R)}})^{\gamma}\nonumber \delta(r)\\
&&+(W+BP_\sigma-HP_{\tau}-MP_{\sigma}P_{\tau})f(r),
\end{eqnarray}
where, $f(r)$ is the functional form of the finite range interaction in term of Gaussian function
$f(r)= e^{-r^2/\alpha^2}$ which contain a single range parameter $\alpha$. The other terms having
their usual meaning \cite{bhuyan12}.\\
The Hamiltonian density functional using simple effective interaction is written as
\begin{eqnarray}
\cal{H}&=& {\cal{K}}+{\cal{H}}^{Nucl}+{\cal{H}}^{SO}(\textbf{r})\nonumber\\
&&+{\cal{H}}^{Coul}(\textbf{r})+{\cal{H}}^{RC}.
\end{eqnarray}
Where ${\cal{K}}=\frac{\hbar^{2}}{2m}(\tau_n+\tau_p)$ is the kinetic energy term with $\tau_n$ and
 $\tau_p$ are the proton and neutron kinetic energy densities of nucleus. The second
term of the Hamiltonian is the nuclear contribution which contain the direct and exchange part.
The direct contribution of nuclear energy density comes from the central part of the effective interaction.
The third term is the spin-orbit interaction written as
\begin{eqnarray}
{\cal{H}}^{SO}(\textbf{r})&=&\frac{-1}{2}W_0 [\rho(\textbf{r})\nabla\textbf{J}+\rho_n(\textbf{r})\nabla \textbf{J}_n \nonumber\\
&&+\rho_p(\textbf{r})\nabla \textbf{J}_p ].
\end{eqnarray}
The fourth terms is due to Coulomb interaction which
contains both direct and exchange terms given by
\begin{eqnarray}
{\cal{H}}^{Coul}(\textbf{r})&=& \frac{1}{2}\int\frac{\rho_p(\acute{\textbf{r}})}{|\textbf{r}-\acute{\textbf{r}}|} d^3 \acute{r}-\frac{3}{4}({\frac{3}{\pi}})^{1/3} \rho_p^{4/3}.
\end{eqnarray}
The last term of the equation arises from the zero range part of the SEI,
which plays the role of residual correlation energy.
\begin{eqnarray}
{\cal{H}}^{RC}&=&\frac{t_0}{4}\int[(1-x_0)[\rho_{n}^2(\textbf{r})+\rho_{p}^2(\textbf{r})]]\nonumber\\
&&+\frac{t_0}{4}\int[(4+2x_0)\rho_n(\textbf{r})\rho_p(\textbf{r})]\nonumber\\
&&+\frac{t_3}{24}\int[(1-x_3)[\rho_{n}^2(\textbf{r})+\rho_{p}^2(\textbf{r})]]\nonumber\\
&&+\frac{t_3}{24}\int[(4+2x_3)\rho_n(\textbf{r})\rho_p(\textbf{r})](\frac{\rho(\textbf{r})}{1+b\rho(\textbf{r})})^\gamma.
\end{eqnarray}
Here $\rho_n$, $\rho_p$, $\rho$, $J_n$, $J_p$ and $J$ are the neutron, proton and total nuclear
and current densities respectively. The 12 parameters $\gamma$, $b$, $t_0$, $t_3$, $x_0$, $x_3$,
$W$, $B$, $H$, $M$, $\alpha$ and $W_0$
are used for the calculation of ground state properties. The detailed procedure of calculations for ground state
properties like binding energy, charge radius, nuclear matter radius etc. and parameters has been in Ref. \cite{bhuyan12}.

\subsection{ Relativistic mean field Lagrangian density}
The relativistic mean field formalism is well documented in Refs.
\cite{boguta1977,pannert1987,ring1996,patra1991,Del2001}.
The basic ingredient of RMF model is the relativistic Lagrangian density for a nucleon-meson many body
system which is defined as \cite{boguta1977,pannert1987,patra1991}
\begin{eqnarray}
{\cal L}&=&\overline{\psi}_i(i\gamma^{\mu}\partial_{\mu}-M)\psi_i+{\frac{1}{2}}\partial^{\mu}\sigma\partial_{\mu}\sigma\nonumber\\
&&-\frac{1}{2}m^{2}_{\sigma}\sigma^{2}-\frac{1}{3}g_2\sigma^3-\frac{1}{4}g_3\sigma^4-g_s\overline{\psi}_i\psi_i\sigma\nonumber\\
&&-{\frac{1}{4}}\Omega^{\mu\nu}\Omega_{\mu\nu}+{1\over{2}}m_{w}^{2}V^{\mu}V_{\mu}\nonumber\\
&&-g_\omega\overline{\psi}_i\gamma^{\mu}\psi_iV_{\mu}-{\frac{1}{4}}\vec{B}^{\mu\nu}.\vec{B}_{\mu\nu}\nonumber\\
&&+{1\over{2}}m_{\rho}^{2}{\vec R^{\mu}}.{\vec{R}_{\mu}}-g_\rho\overline{\psi}_i\gamma^{\mu}\overrightarrow{\tau}\psi_i.\overrightarrow{R^{\mu}}\nonumber\\
&&-{1\over{4}}F^{\mu\nu}F_{\mu\nu}-e\overline{\psi}_i\gamma^{\mu}\frac{(1-\tau_{3i})}{2}\psi_iA_{\mu}.
\end{eqnarray}
Here $\sigma$, $V_{\mu}$ and $\overrightarrow{R}_{\mu}$ are the fields for $\sigma$-,
 $\omega$- and $\rho$-meson respectively. $A^{\mu}$ is the electromagnetic
field. The $\psi_i$ are the Dirac spinors for the nucleons whose third
component of isospin is denoted by $\tau_{3i}$. $g_s$, $g_\omega$,
$g_\rho$ and $\frac{e^2}{4\pi}=\frac{1}{137}$ are the coupling constants
for the linear term of $\sigma$-, $\omega$- and $\rho$-mesons and photons
respectively. $g_2$ and $g_3$ are the parameters for the non-linear
terms of the $\sigma$-meson.
M, $m_\sigma$, $m_\omega$ and $m_\rho$ are the masses of the nucleons,
$\sigma$-, $\omega$- and $\rho$-mesons, respectively. $\Omega^{\mu\nu}$,
$\overrightarrow{B}^{\mu\nu}$ and $F^{\mu\nu}$ are the field tensors for
the $V^{\mu}$, $\overrightarrow{R}^{\mu}$ and the photon fields, respectively.
Now we used two assumptions, first the nucleons are moving inside
the nucleus in spherically symmetric potential. In this case, the large and small component of
Dirac spinor $\psi_i$ are expanded separately in terms of radial function of a spherical harmonic
oscillator potential and in second assumption, the nucleons
are moving in non spherical symmetric potential. In this case the large and small component of
Dirac spinor are expanded in axial symmetric manner in term of deformed harmonic oscillator potential
by taking volume conservation into account. The set of Dirac and Klein-Gorden equation are
solved by these two assumptions to obtain the bulk properties of nuclei.
The quadrupole moment deformation parameter ($\beta_2$), root mean square radii ($r_m$)
and binding energy (B.E.) are evaluated using the standard relations \cite{patra1991}.

\subsection{Glauber Model}
\subsubsection{(i) Reaction cross section}
The theoretical formalism to study the reaction cross sections
using the Glauber approach has been given by R. J. Glauber \cite{glauber1955,glau55}.
The nucleus-nucleus elastic scattering amplitude is
written as
\begin{equation}
F(q)=\frac{\iota K}{2\pi} \int db e^{\iota q.b}(1-e^{\iota \chi(b)}).
\end{equation}
At low energy,  this model is modified in order to take care of finite
range effects in the profile function and
Coulomb modified trajectories. The elastic scattering amplitude including
the Coulomb interaction is expressed as
\begin{equation}
F(q)=e^{\iota \chi_{s}}\{F_{coul}(q)+\frac{\iota K}{2\pi}
\int db e^{\iota q.b+2\iota \eta \ln(Kb)}(1-e^{\iota \chi(b)})\},
\end{equation}
with the Coulomb elastic scattering amplitude
\begin{equation}
F_{coul}(q)=\frac{-2 \eta K}{q^2}exp\{-2 \iota \eta \ln(\frac{q}{2K})
+2\iota arg \Gamma(1+\iota \eta)\},
\end{equation}
where $K$ is the momentum of projectile and $q$ is the momentum transferred
from the projectile to the target. Here $\eta=Z_P Z_T e^2/\hbar v$ is the
Sommerfeld parameter, $v$ is the incident velocity of the projectile,
and $\chi_s=-2\eta \ln(2Ka)$ with $a$ being a screening radius.
The elastic differential cross section is given by
\begin{equation}
\frac{d\sigma}{d\Omega}=|F(q)|^2.
\end{equation}
The standard Glauber form for total reaction cross
sections is expressed as \cite{glau55,Karo75}
\begin{equation}
\sigma_R=2\pi\int_0^\infty b[1-T(b)]db,
\end{equation}
where 'T(b)' is the Transparency function with impact parameter
'b'. The function T(b) is calculated by
\begin{equation}
T(b)=exp[-\sum_{i,j}\sigma_{ij}\int\overline{\rho_{tj}}(s)
\overline{\rho_{pi}}(|\overrightarrow{b}-\overrightarrow{s}|)
\overrightarrow{ds}].
\end{equation}
Here, the summation indices i, j run over proton and neutron and
subscript 'p' and 't' refers to projectile and target
respectively. $\sigma_{ij}$ is the experimental nucleon-nucleon
reaction cross-section which depends on the energy. The z-
integrated densities are defined as
\begin{equation}
\overline{\rho}(\omega) =\int_{-\infty}
^\infty\rho(\sqrt{w^{2}+z^{2}}) dz,
\end{equation}
with $\omega^2=x^2+y^2$.
Initially Glauber model was designed for the
high energy approximation. However it was found to work reasonably
well for both the nucleus-nucleus reaction and the differential
elastic cross-sections
over a broad energy range \cite{Chauvin1983,Buenerd1984}.
The modified transparency function T(b) is given by
\begin{equation}
 T(b)=exp[-\int_p\int_t\sum_{i,j}[\Gamma_{ij}(\overrightarrow{b}-\overrightarrow{s}+\overrightarrow{t})]
 \overrightarrow{\rho_{pi}}(\overrightarrow{t})
\overrightarrow{\rho_{tj}}(\overrightarrow{s})
\overrightarrow{ds}\overrightarrow{dt}].
\end{equation}
The profile function $\Gamma_{NN}$ for optical limit approximation is
defined as
\begin{equation}
\Gamma_{NN} =\Gamma_{ij}(b_{eff}) =\frac{1-\iota\alpha_{NN}}{2\pi
\beta^2_{NN}}\sigma_{NN} exp(-\frac{b^2_{eff}}{2\beta^2_{NN}}),
\end{equation}
for finite range and
\begin{equation}
\Gamma_{NN} =\Gamma_{ij}(b_{eff}) =\frac{1-\iota\alpha_{NN}}{2}\sigma_{NN} \delta (b),
\end{equation}
for zero range
with
$b_{eff}$=$|\overrightarrow{b}-\overrightarrow{s}+\overrightarrow{t}|$, $\overrightarrow{b}$ is the impact parameter. Where
$\overrightarrow{s}$ and $\overrightarrow{t}$ are the dummy variables for integration
over the z-integrated target and projectile densities. The
parameters $\sigma_{NN},\alpha_{NN}$ and $ \beta_{NN}$ are usually
depends upon the proton-proton, neutron-neutron and proton-neutron
interactions.

\section{Calculations and results}
\subsection{Ground state properties}
\begin{table*}
\renewcommand{\arraystretch}{1.0}
\tabcolsep 0.08cm
\caption{\label{tab:table1} The ground state properties of projectile and
target nuclei obtained from RMF(NL3) and HF(SEI) calculations are
compared with experimental data wherever available. The binding energy (B.E.) is in MeV and
charge radius $r_c$ is in fm.}
\begin{tabular}{lcccccccccccr}
\hline
\hline
\multicolumn{3}{r}{B.E.}&\multicolumn{5}{r}{$r_c$} & \multicolumn{4}{r}{$\beta_2$} \\
\cline{2-5} \cline{7-10} \cline{12-13}
  Nuclei&HF(SEI)&RMF(NL3)&RMF(NL3) &Expt.\cite{audi2003,priv2012}&&HF(SEI)&RMF(NL3)&RMF(NL3)&Expt.\cite{angeli13}&& RMF(NL3)&Expt.\cite{data}\\
&(Sph.)&(Sph.)&(Def.)&&&(Sph.)&(Sph.)&(Def.)&&&(Def.)\\
\hline\hline \\
 $^{9}Be$&54.927 &54.76&58.018&58.164$\pm$0.000&&2.305&2.461&2.510&2.519&&0.793&  \\
 $^{10}Be$&65.302 &63.49&64.855 &64.970$\pm$0.000&&2.302&2.537&2.423&2.36&&0.509&  \\
 $^{11}Be$&69.380 &67.97&67.780 &65.478$\pm$0.000&&2.329&2.479&2.449&2.46&&0.369&  \\
 $^{12}Be$&73.190 &73.61&71.378 &68.649$\pm$0.004&&2.353&2.549&2.450&&&-0.137&  \\
 $^{12}B$&82.172 &82.85&82.176 &79.575$\pm$0.001&&2.393&2.498&2.457&&& 0.168&  \\
 $^{13}B$&88.250 &88.84&88.842&84.453$\pm$0.001&&2.411&2.540& 2.492&&&0.097& \\
 $^{14}B$&89.256 &91.88&89.874 &85.423$\pm$0.021&&2.423&2.534&2.522&&&0.382&  \\
 $^{15}B$&90.162 &93.64&92.593 &88.194$\pm$0.022&&2.435&2.532&2.564&2.511&&0.611& \\
 $^{12}C$&88.422&88.23&91.349&92.160$\pm$1.700&&2.436&2.364&2.310 &2.47&&0.007& 0.577(16)\\
 $^{13}C$&97.489 &96.31&98.098&97.108$\pm$0.000&&2.454&2.459&2.466&2.46&&-0.000&  \\
 $^{14}C$&105.829 &104.38&106.929 &105.284$\pm$0.000&&2.470&2.504&2.517&2.56&&0.000& 0.36(3)\\
 $^{15}C$&108.846 &108.66&108.477 &106.502$\pm$0.000&&2.480&2.516&2.535&&&0.249 & \\
 $^{16}C$&111.642 &113.45&111.874 &110.752$\pm$0.003&&2.489&2.531&2.565&&&0.448 & \\
 $^{17}C$&114.248 &116.40&114.083&111.486$\pm$0.017&&2.500&2.542&2.582&&&0.458&\\
 $^{18}C$&116.709 &118.79&116.842 &115.670$\pm$0.030&&2.509&2.552&2.601&&&0.471&\\
 $^{19}C$&119.015 &119.91&119.511& 116.242$\pm$0.100&&2.521&2.562&2.629&&&-0.438&\\
 $^{20}C$&121.154 &122.54&119.736&119.18$\pm$0.200 &&2.532&2.573&2.590&&&0.278&\\
 $^{20}N$&139.487 &139.37&136.772&134.184$\pm$0.055&&2.605&2.671&2.683&&&-0.306&\\
 $^{21}N$&143.095 &142.59&141.475&138.768$\pm$0.100&&2.614&2.674&2.694&&&-0.311&\\
 $^{22}N$&146.205 &145.55&143.948&140.052$\pm$0.200&&2.624&2.679&2.678&&&-0.159&\\
 $^{23}N$&148.765 &148.17&147.864&141.726$\pm$0.300&&2.637&2.687&2.674&&&-0.009&\\
 $^{20}O$&154.043 &152.71&151.585&151.36$\pm$0.100&&2.668&2.720&2.726&&&0.250 & 0.268(6)\\
 $^{21}O$&159.645 &157.95&156.944&158.928$\pm$0.100&&2.673&2.721&2.716&&&0.132&\\
 $^{22}O$&164.758 &164.18&163.192&166.496$\pm$0.100&&2.678&2.735&2.712&&&0.002&0.19(4)\\
 $^{23}O$&169.040 &168.32&167.227&174.064$\pm$0.100&&2.689&2.741&2.725&&&0.003&\\
 $^{24}O$&172.508 &171.87&171.665&181.632$\pm$0.100&&2.701&2.752&2.737&&&0.003&\\
 $^{23}F$&177.956 &176.75&175.378&175.283$\pm$0.100&&2.768&2.839&2.805&&&-0.188&\\
 $^{24}F$&183.433 &182.06&180.162&179.11$\pm$0.100&&2.778&2.838&2.812&&&-0.128&\\
 $^{25}F$&186.026 &186.90&185.221&183.375$\pm$0.100&&2.782&2.853&2.821&&&-0.087&\\
 $^{26}F$&188.890 &191.76&187.928&184.158$\pm$0.100&&2.806&2.875&2.852&&&-0.125&\\
 $^{27}F$&191.507 &195.00&191.245&186.246$\pm$0.200&&2.827&2.891&2.886&&& 0.151&\\
 $^{28}Ne$&207.273&210.62&208.122&206.89$\pm$ 0.100&&2.892&2.964&2.966&2.963&&0.223&0.36(3)\\
 $^{29}Ne$&210.833&214.35&211.140 &207.843$\pm$0.100 &&2.912&2.982&2.981&&&0.159& \\
 $^{30}Ne$&214.160&218.02&214.920&211.29$\pm$0.300&&2.933&2.998&2.999&&&0.098&0.49(17)\\
 $^{31}Ne$&214.569&221.97&215.812&211.42$\pm$0.200&&2.944&3.012&3.032&&&0.238 &\\
 $^{32}Ne$&214.860&223.95&218.409&213.472$\pm$0.500&&2.956&3.024&3.071&&&0.363&\\
 $^{32}Mg$&249.390&252.06&250.387 &249.804$\pm$0.200&&3.032&3.095&3.091&3.186&&0.119& 0.51(5)\\
 $^{33}Mg$&252.015&256.02&252.982&252.017$\pm$0.200&&3.043&3.107&3.118&&&0.231 &\\
 $^{34}Mg$&254.355&259.14&257.169&256.462$\pm$0.100&&3.053&3.118&3.151&&&0.340& 0.55(6)\\
 $^{35}Mg$&256.498&261.82&260.211&257.460$\pm$0.200&&3.064&3.129&3.174&&&0.385&\\
 $^{32}Si$&267.928&267.69&268.203&271.407$\pm$0.000&&3.078&3.113&3.141&&&-0.203& 0.26(4)\\
 $^{33}Si$&275.953&275.97&275.359&275.915$\pm$0.000&&3.095&3.133&3.134&&&-0.085&\\
 $^{34}Si$&283.470&283.78&278.249&283.428$\pm$0.014&&3.111&3.153&3.206&&&-0.337& 0.18(4)\\
 $^{35}Si$&288.258&289.77&287.135 &285.903$\pm$0.038&&3.119&3.163&3.164&&&-0.084&\\
 $^{34}S$&286.424&286.05&286.295 &291.838$\pm$0.000&&3.199&3.270&3.259&3.28&&-0.168& 0.247(3)\\
 $^{35}S$&296.491&296.71&295.543&298.824$\pm$0.000&&3.209&3.282&3.262&&&-0.077&\\
 $^{36}S$&305.978&306.52&299.502&308.714$\pm$0.000&&3.219&3.293&3.312&3.29&&-0.309& 0.157(7)\\
 $^{37}S$&312.676&313.58&309.946&313.017$\pm$0.000&&3.224&3.299&3.290&&&-0.116&\\
 $^{34}Ar$&273.357&273.76&273.397&278.719$\pm$0.000&&3.295&3.387&3.360&3.365&&-0.168&0.229(15)\\
 $^{36}Ar$&300.129&300.25&302.268&306.716$\pm$0.000&&3.305&3.388&3.379&3.390&&-0.209&0.253(8)\\
 $^{38}Ar$&324.432&325.59&319.946&327.342$\pm$0.000&&3.318&3.397&3.396&3.402&&-0.279&0.161(4)\\
 $^{40}Ar$&341.555&342.31&340.945&343.810$\pm$0.000&&3.326&3.401&3.392&3.427&& -0.160&0.269(5)\\
 $^{42}Ar$&357.013&357.39&356.393&359.335$\pm$0.005&&3.335&3.406&3.402&3.435&&-0.176&0.27(3)\\
 $^{44}Ar$&371.135&371.42&370.639&373.728$\pm$0.001&&3.345&3.410&3.410&3.445&&-0.179&0.22(16)\\
 $^{46}Ar$&383.672&384.57&384.603&386.927$\pm$0.040&&3.373&3.410&3.415&3.437&&-0.167&0.170(17)\\
 $^{48}Ar$&391.79&394.58&392.810&396$\pm$0.720&&3.393&3.437&3.442&&&-0.198&\\
 \hline\hline
 \end{tabular}
 \end{table*}

The RMF(NL3) and HF(SEI) explain the nuclear properties like B.E. and
rms radii reasonably well for almost all nuclei in the periodic chart.
We use these sets to discuss the bulk properties of the considered nuclei.
\begin{figure}
\includegraphics[width=1.0\columnwidth,clip=true]{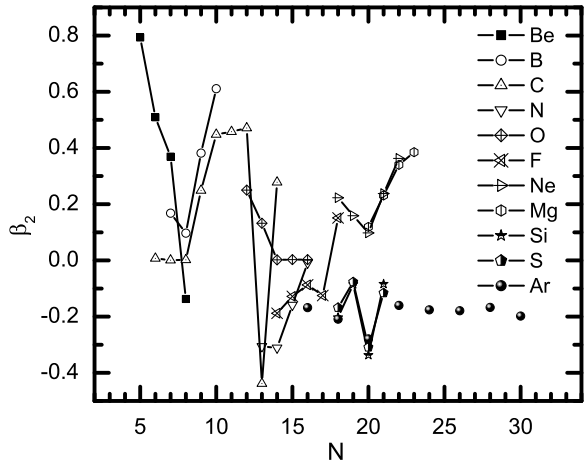}
\caption{The quadrupole deformation parameter $\beta_2$ of $^{9-12}$Be, $^{12-15}$B, $^{12-20}$C, $^{20-23}$N, $^{20-24}$O,
$^{23-27}$F, $^{28-32}$Ne, $^{32-35}$Mg, $^{32-35}$Si, $^{34-37}$S and $^{34-48}$Ar  as a function of
neutron number (N) for RMF(NL3).} \label{Fig.1}
\end{figure}

\subsubsection{Binding energy}
The binding energies (B.E.'s) of $^{9-12}$Be, $^{12-15}$B, $^{12-20}$C, $^{20-23}$N, $^{20-24}$O,
$^{23-27}$F, $^{28-32}$Ne, $^{32-35}$Mg, $^{32-35}$Si, $^{34-37}$S and $^{34-48}$Ar obtained by
relativistic mean field theory using spherical and axially deformed coordinates systems and with non relativistic mean
field theory using simple effective interaction are presented in
Table I along with experimental values. The B.E. of $^{10}$Be are 65.302, 63.49 and 64.855 MeV for HF(SEI), Sph. RMF(NL3)
and Def. RMF(NL3) formalisms, well comparable to its experimental value 64.970 MeV. Hence by examining the results
of Table I, it can be concluded that both the formalisms are capable
to reproduce the experimental data for the considered nuclei.

\subsubsection{Charge radius}
\begin{figure*}
\includegraphics[width=2.0\columnwidth,clip=true]{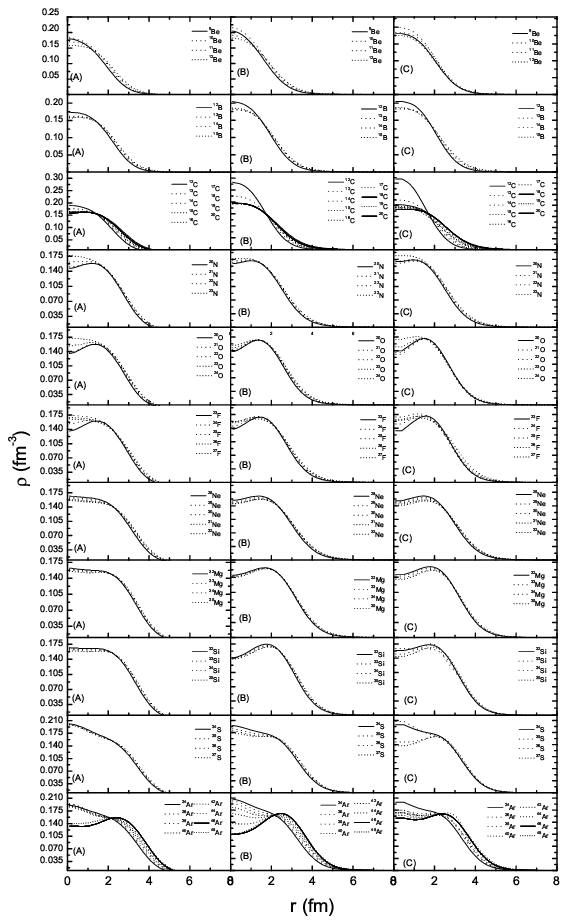}
\vspace{-1.0cm}
\caption{ Radial density plots for various isotopes of $^{9-12}$Be, $^{12-15}$B, $^{13-20}$C, $^{20-23}$N, $^{20-24}$O,
$^{23-27}$F, $^{28-32}$Ne, $^{32-35}$Mg, $^{32-35}$Si, $^{34-37}$S and $^{34-48}$Ar obtained by (A) HF(SEI)
(B) Spherical RMF(NL3) (C)Deformed RMF(NL3) formalisms.} \label{Fig.1}
\end{figure*}

The calculated root mean square charge radius ($r_c$) of projectile
and target nuclei from both RMF(NL3) and non
relativistic HF(SEI-I) mean field theories are presented in Table I.
The experimental data are also given for comparison wherever
available \cite{angeli13}. The rms proton radius $r_p$ is obtained from
the distribution of point protons inside the nucleus. The charge radius $r_c$ is calculated by taking
the finite size 0.8 fm of the proton, which is evaluated from the formula
$r_c=\sqrt{r^{2}_p+0.06}$ \cite{patra1991}. The calculated values of $r_c$ for $^{11}$Be are 2.329, 2.479 and 2.449 fm from HF (SEI), Sph.
RMF(NL3) and Def. RMF(NL3) formalisms, which are well comparable with the experimental value of 2.46 fm. Similarly
$r_c$ for $^{36}$Ar and $^{46}$Ar are 3.035 fm, 3.373 fm for HF(SEI), 3.388 fm, 3.410 fm for Sph. RMF(Nl3), 3.379 fm, 3.415 fm for
Def. RMF(NL3) and 3.390 fm, 3.437 fm for experimental observations, respectively. In general, observation from this table
signifies the successability of these theories by predicting the surprisingly comparable results with experimental data.

\begin{table*}
\renewcommand{\arraystretch}{1.0}
\tabcolsep 0.20cm
\caption{\label{tab:table1} The Depletion Factor (D.F in \%) of neutron (N), proton (P) and total (T) densities for
some probable cases of bubble nuclei obtained from HF(SEI), Sph. RMF(NL3) and Def. RMF(NL3).}
\begin{tabular}{lcccccccccccr}
\hline\hline
\multicolumn{3}{r}{D.F\%}& \multicolumn{4}{r}{D.F\%} & \multicolumn{4}{r}{D.F.\%} \\
\cline{2-4} \cline{6-8} \cline{10-12}
 Nuclei&& HF(SEI)&&&&Sph. RMF&&&& Def. RMF&\\
&N&P&T&&N&P&T&&N&P&T\\
\hline\hline \\
 $^{22}$O& 8.21&19.14&13.21&&9.49&20.23&14.46&& 24.35&22.47&23.29\\
 $^{23}$F& 10.67&20.45&15&&14.44&22.71&18.26&&22.14&22.30&21.99\\
 $^{34}$Si&-&36.66&-&&0.36&36.27&16.75&&-&3.73&-\\
 $^{36}$S&-&-&- &&-&-&-&&3.49&16.25&9.49 \\
 $^{36}$Ar&-&-&-&&-&-&-&&-&-&-\\
 $^{46}$Ar&-&51.05&14.74&&15.31&62.08&36.64&&1.56&14.73&7.51 \\
 \hline\hline
 \end{tabular}
 \end{table*}

\begin{figure}
\hspace{-0.7cm}
\includegraphics[width=1.1\columnwidth,clip=true]{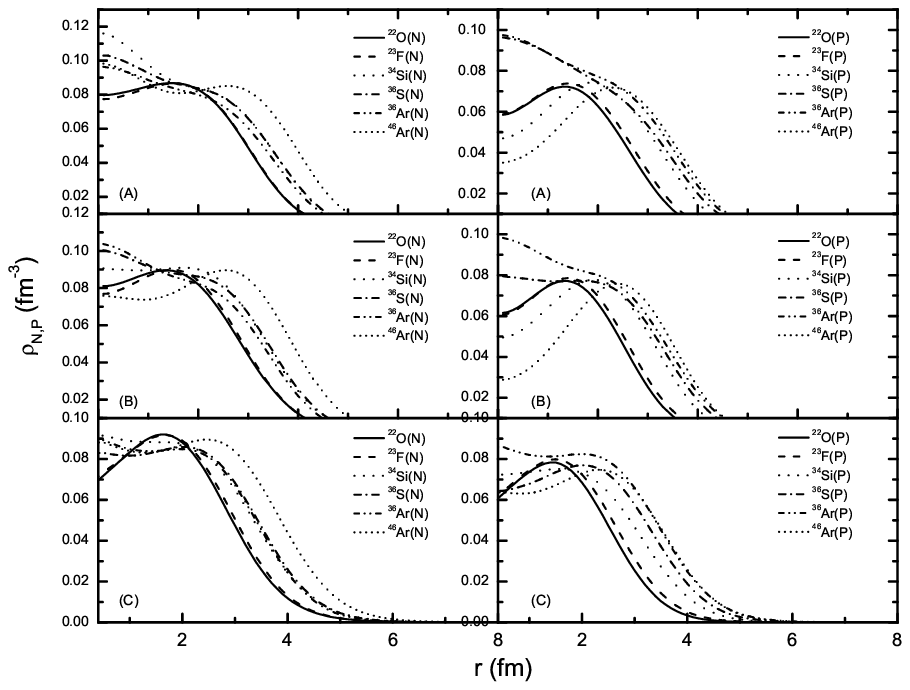}
\caption{Radial density plots for expected bubble nuclei
$^{22}$O, $^{23}$F, $^{34}$Si, $^{36}$S $^{36}$Ar and $^{46}$Ar obtained by (A) HF(SEI)
(B) Sph. RMF(NL3) (C) Def. RMF(NL3) formalisms.} \label{Fig.1}
\end{figure}

\subsubsection{Quadrupole deformation parameter $ \beta_2$}
Figure 1 shows the quadrupole deformation parameter ($\beta_2$) of considered cases of nuclei obtained from RMF(NL3) formalism
as a function of their neutron numbers of the isotopes.
Table I also presents the values of deformation along with the experimental values which are available.
The negative $\beta_2$ values of nuclei in Table I signifies its oblate deformation where as
positive values of prolate deformation and zero value of spherical nature. It is worthy to
mention that the NL3 parameter set does not give a converged solution for many of the light
nuclei as such. To get a converged result for such cases, we change the pairing strength
$\triangle_{n,p}$ slightly in the BCS-pairing approach. As a result, we compromise a bit in
the quadrupole deformation.

\subsubsection{Density}
Fig. 2 presents the densities of the considered nuclei as a function of radial distance (r in fm).
The nucleons distribution inside the nucleus is maximum at the centre and starts decreasing continuously towards the surface.
The left panel of the figure presents the nucleonic density distribution obtained by non relativistic mean field
HF(SEI) approach. The right panel shows spherical equivalent of deformed RMF densities
with NL3 parameter. The central panel of the figure shows the density distribution obtained from spherical
symmetric RMF model. It is clear from the figure that the densities of considered nuclei shows similar kind of trend
from all the formalisms. A deep inspection of the figure indicates some of the isotopes of O, F, Si, S and Ar
shows depletion of the densities at the centre, which is the primary indication for their bubble structure.\\

Some of the prominent cases of bubble nuclei are plotted in Fig. 3. This figure contains the proton and neutron density
distributions of $^{22}$O, $^{23}$F, $^{34}$Si, $^{36}$S $^{36}$Ar and $^{46}$Ar nuclei as a function of radial distance.
The left panel of the figure shows the neutron  and right panel shows
proton density distribution of considered cases of nuclei for HF(SEI), Sph. RMF(NL3) and
Def. RMF(NL3) densities, respectively. It is clear from the figure, the depletion for proton density distribution
is more than the neutron counterpart. Detailed
discussion of bubble nuclei has been given in next subsection.

\begin{table*}
\renewcommand{\arraystretch}{1.2}
\tabcolsep 0.05cm
\caption{\label{tab:table3}The Gaussian coefficients $c_1$, $a_1$, $c_2$, $a_2$ of projectile
and target for HF(SEI), Sph. RMF(NL3) and Def. RMF(NL3) densities.}
\begin{tabular}{ccccccccccccccccc}
\hline
\multicolumn{4}{r}{HF(SEI-I)}& \multicolumn{5}{r}{Sph. RMF(NL3)}& \multicolumn{5}{r}{Def. RMF(NL3)} \\
\cline {2-5} \cline{7-10} \cline {12-15}
$Nuclei$& $c_1$& $a_1$ &$c_2$ & $ a_2$&& $c_1$& $a_1$ &$c_2$ & $ a_2$&& $c_1$&$a_1$& $c_2$&$ a_2$\\
\hline
\hline
 $^{9}Be$ & -2.51177&0.393812&2.69169&0.37377 &&-0.0593744&0.659468&0.268817&0.289419& &-1.25421& 0.39907 & 1.39728 & 0.355553 \\
 $^{10}Be$ &-3.33309&0.379659&3.51516&0.363426 &&-0.0915913&0.609445&0.312021&0.291001&& -1.22148& 0.413105&1.39161  &0.361781  \\
 $^{11}Be$ &-3.04829&0.355441&3.2177&0.338208 &&-0.122702&0.527202&0.319249&0.268455& &-1.21881 & 0.390577 & 1.38332 & 0.340326 \\
 $^{12}Be$ &-1.35748&0.35003&1.51497&0.311617  &&-0.311776&0.455307&0.521176&0.300176&&-0.995759 & 0.218859 & 1.20496 &0.217494 \\
 $^{12}B$ &-3.63908&0.350342&3.80913&0.334343&&-1.50466&0.407157&1.70762&0.358788&&-1.19869&0.404408&1.39307&0.347786\\
 $^{13}B$ &-3.94932&0.335534&4.10771&0.320364&&-2.67497&0.375355&2.85669&0.347163&&-0.499802&0.428431&0.677544&0.30021\\
 $^{14}B$ &-3.93077&0.321021&4.08801&0.306225&&-0.86619&0.390135&1.04795&0.310126&&-0.210262&0.478779&0.376203&0.242406\\
 $^{15}B$ &-1.60867&0.317723&1.76563&0.283852&&-0.466283&0.412755&0.645633&0.276104&&-1.07423&0.297112&1.23432&0.259716\\
 $^{12}C$ &-3.79616&0.361674&3.98071&0.345423& &-0.232654&0.638687&0.0.517232&0.339911&&-1.14333&0.285974&1.47032&0.285598 \\
 $^{13}C$&-1.74891&0.355312&1.9196&0.320299&&-1.44407&0.413914&1.64276&0.355309&&-1.17196&0.403733&1.40401&0.345876\\
 $^{14}C$&-1.88677&0.342428&2.04598&0.308693&&-3.77882&0.365346&3.96791&0.344481&&-1.18651&0.385936&1.37994&0.324587\\
 $^{15}C$&-1.90963&0.327595&2.0674&0.295423&&-4.16046&0.35900&4.33524&0.338276&&-0.294377&0.45885&0.472115&0.256238\\
 $^{16}C$&-1.9147&0.314974&2.07175&0.283792&&-1.6802&0.35973&1.85217&0.311381&&-1.21984&0.316502&1.38936&0.27335\\
 $^{17}C$&-1.91276&0.303614&2.06889&0.273238&&-3.52531&0.332303&3.69421&0.3095&&-1.15816&0.309391&1.32105&0.263253\\
 $^{18}C$&-1.93748&0.292373&2.0941&0.263313&&-0.22627&0.429346&0.432019&0.225458&&-1.19721&0.300572&1.35334&0.255236\\
 $^{19}C$&-1.94062&0.280481&2.09825&0.25306&&-2.6298&0.312787&2.79198&0.284036&&-0.61523&0.331571&0.866337&0.257346\\
 $^{20}C$&-4.27647&0.261466&4.43707&0.24962&&-2.58028&0.302332&2.73839&0.273816&&-0.162047&0.423205&0.350278&0.191529\\
 $^{20}N$&-2.27523&0.279519&2.41666&0.253036&&-1.65348&0.298831&1.82293&0.260481&&-0.760729&0.309713&0.966724&0.24428\\
 $^{21}N$&-2.28047&0.269391&2.42488&0.244319&&-1.28808&0.292176&1.46044&0.246603&&-0.795721&0.306924&0.998833&0.240861\\
 $^{22}N$&-2.16896&0.25557&2.32419&0.232462&&-1.21886&0.280349&1.39582&0.236182&&-1.16072&0.24254&1.36471&0.215256\\
 $^{23}N$&-4.08043&0.234574&4.25039&0.224176&&-1.21542&0.266098&1.39836&0.226368&&-0.329494&0.133146&0.523324&0.132026\\
 $^{20}O$ &-2.47486&0.286301&2.60404&0.25909 &&-5.45193&0.299258&5.58882&0.284173&&-1.93556&0.310064&2.05253&0.267917 \\
 $^{21}O$ &-2.53565&0.27829&2.66435&0.252158 &&-2.00566&0.303822&2.14271&0.264553&&-1.51258&0.315625&1.63101&0.260035\\
 $^{22}O$ &-2.57536&0.269317&2.7064&0.244573 &&-1.86978&0.293242&2.01341&0.254045&&-0.64512&0.363379&0.764856&0.230878 \\
 $^{23}O$ &-2.43928&0.254471&2.58493&0.232008 &&-1.80269&0.282707&1.9515&0.244704&&-0.657758&0.330368&0.802805&0.223554\\
 $^{24}O$ &-2.04609&0.239581&2.20933&0.217663&&-1.92109&0.267916&2.07684&0.235596&&-0.668732&0.297564&0.838554&0.214713\\
 $^{23}F$ &-2.69853&0.263769&2.82652&0.23994 &&-5.77017&0.27426&5.90555&0.260946&&-2.48442&0.288719&2.59277&0.254931\\
 $^{24}F$ &-2.42788&0.249261&2.57181&0.226755 &&-2.13084&0.274675&2.27389&0.241917&&-0.885032&0.274091&1.06166&0.217354\\
 $^{25}F$ &-2.09825&0.2342&2.26064&0.212972 &&-1.99848&0.261501&2.15224&0.230625&&-0.763297&0.238881&0.956929&0.196001\\
 $^{26}F$ &-2.06675&0.224826&2.22373&0.20445 &&-1.94522&0.249237&2.09993&0.220358&&-0.742259&0.240026&0.928872&0.19271\\
 $^{27}F$ &-2.03263&0.216267&2.18559&0.196679&&-1.94413&0.238163&2.09571&0.211318&&-1.85421&0.23814&1.99459&0.209488 \\
 $^{28}Ne$ &-2.10042&0.212881&2.2517&0.19371 &&-2.17488&0.232385&2.32326&0.208284&&-0.937098&0.256729&1.11394&0.202448\\
 $^{29}Ne$ &-2.07765&0.205597&2.22506&0.187105 &&-2.31443&0.224436&2.45925&0.202571&&-0.919743&0.232073&1.09489&0.187997\\
 $^{30}Ne$ &-2.05061&0.198501&2.19404&0.180662 &&-2.34129&0.215415&2.48218&0.195104&&-0.815071&0.213394&0.985336&0.173332\\
 $^{31}Ne$ &-2.07274&0.194286&2.2139&0.176848&&-2.30031&0.214514&2.44391&0.193464&&-0.896996&0.216244&1.0655&0.17556 \\
 $^{32}Ne$ &-4.57748&0.184948&4.71668&0.177083 &&-2.26169&0.211097&2.40614&0.189954&&-0.928504&0.2197&1.09451&0.176494\\
 $^{32}Mg$ &-2.20212&0.194812&2.34336&0.177429&&-2.60897&0.212996&2.74249&0.193239&& -0.967166&0.213458&1.13137&0.174275\\
 $^{33}Mg$ &-2.22698&0.191053&2.36594&0.174019&&-2.60291&0.209569&2.73607&0.18993&& -1.07939&0.221034&1.24224&0.179671\\
 $^{34}Mg$ & -2.24103&0.18709&2.37778&0.170425&&-2.58898&0.206733&2.72212&0.187059&&-1.06206&0.23098&1.22155&0.182458\\
 $^{35}Mg$ &-2.26351&0.183622&2.39807&0.167282&&-2.5666&0.203774&2.69959&0.184066&& -1.05435&0.230207&1.21082&0.180174\\
 $^{32}Si$ &-2.37277&0.202381&2.51805&0.184376&&-3.00897&0.227291&3.13661&0.206144&& -0.975703&0.260388&1.13915&0.19639\\
 $^{33}Si$ &-2.35386&0.196639&2.49616&0.179151&&-2.9272&0.218848&3.05298&0.198521&& -0.995642&0.223719&1.1529&0.178138\\
 $^{34}Si$ & -2.34203&0.191388&2.48123&0.174374&&-2.86769&0.211811&2.99191&0.192149&&-1.25717&0.26293&1.42724&0.204504\\
 $^{35}Si$ & -2.36933&0.187986&2.50617&0.171288&&-2.88628&0.208281&3.00941&0.188926&&-1.09127&0.208147&1.24501&0.169924\\
 $^{34}S$ &-1.81296&0.181829&1.98726&0.165612&&-2.03669&0.193789&2.20712&0.175664&& -1.18862&0.215201&1.38091&0.180478\\
 $^{35}S$ & -1.78947&0.176802&1.96139&0.161028&&-2.11099&0.189798&2.27085&0.172077&&-1.18643&0.175455&1.38282&0.156079\\
 $^{36}S$ & -1.76837&0.172069&1.93796&0.156714&&-2.13714&0.185085&2.28947&0.167819&&-1.52688&0.248852&1.67126&0.199451\\
 $^{37}S$ & -1.79518&0.169348&1.96303&0.154256&&-2.21565&0.183299&2.3634&0.166219&&-1.07227&0.2135&1.21247&0.168601\\
 $^{34}Ar$&-1.93044&0.183573&2.09551&0.167211&&-1.84812&0.18999&2.02988&0.172144&&-1.18677&0.213069&1.38036&0.179369\\
 $^{36}Ar$&-1.83132&0.172884&1.99595&0.157479&&-1.79242&0.177609&1.96319&0.160928&&-1.25018&0.223959&1.41575&0.181827\\
 $^{38}Ar$&-1.79848&0.164912&1.96024&0.150194&&-1.82537&0.169399&1.98489&0.153507&&-1.76793&0.225418&1.92718&0.189616\\
 $^{40}Ar$&-1.85865&0.160185&2.01657&0.145936&&-1.99749&0.166945&2.14665&0.151332&&-1.18469&0.194597&1.34956&0.160534\\
 $^{42}Ar$&-1.91109&0.155909&2.06565&0.142049&&-2.26358&0.166337&2.39771&0.150869&&-1.18071&0.197867&1.33743&0.1597\\
$^{44}Ar$&-1.96876&0.151999&2.11994&0.138517&&-2.68801&0.168372&2.79633&0.152825&&-1.18782&0.195138&1.33648&0.156048\\
 $^{46}Ar$&-2.71167&0.158333&2.8162&0.144599&&-3.13997&0.170354&3.22018&0.154714&&-1.16205&0.189227&1.29828&0.149873\\
 $^{48}Ar$&-2.68582&0.155294&2.80113&0.141718&&-3.15705&0.167492&3.24426&0.152135&&-1.19557&0.179137&1.35593&0.145673\\
 \hline
 \hline
\end{tabular}
\end{table*}

\subsection{Bubble nuclei}
The bubble effect is appeared in some of the nuclei, where the density of nucleus is depleted at
central part. The main mechanism for the formation of bubble
nuclei is the lack of particles in the centre of nucleus which causes the $s$ levels to be less bound than observed
in the usual cases with the uniform density distribution. If the particles rise high enough in energy, highest
$s$ levels will be empty, hence depletion of central density of particles takes place as a consequence, the
radius of the nucleus increases.
The depleted density of nuclei has been measured in term of depletion factor (D. F.) defined as \cite{grasso2009}
 \begin{equation}
 D.F. =\frac{\rho_{max}-\rho_{cen}}{\rho_{max}},
 \end{equation}
where $\rho_{max}$ and $\rho_{cen}$ represent the values of maximum and the central density.
 The calculated values of D. F. in \% for the $^{22}$O, $^{23}$F, $^{34}$Si, $^{36}$S $^{36}$Ar and $^{46}$Ar are
presented in Table II.
 The value of $(D.F)_T$ in \% for $^{22}$O are 13.21\%, 14.46\% and 23.28\% for the HF(SEI), Sph. RMF(NL3) and Def.
RMF(NL3) densities. Similarly the $(D. F.)_T$ for the $^{23}$F are 15\%, 18.26\% and 21.99\% from same densities, respectively.
The ($D.F.)_P$ in \% for the $^{34}$Si and $^{46}$Ar nuclei are 36.36\% and 51.06\% for HF(SEI), 36.27\% and 62.08\% for Sph.
RMF(NL3) and 3.72\% and 14.72\% for Def. RMF(NL3) densities indicate the nature of their proton bubble. The observations from
this table also signifies, no bubble effect for $^{36}$Ar and 16.25\% $(D.F.)_P$ for $^{36}$S in Def. RMF(NL3) case indicates,
it may be a case for proton bubble along with $^{34}$Si and $^{46}$Ar nuclei. Thus prominent cases for bubble effects are found
in $^{22}$O, $^{23}$F, $^{34}$Si and $^{46}$Ar from our study.

\begin{table}
\begin{center}
\caption{\label{tab:table6}The nucleon-nucleon cross-section
$\sigma_{NN}$ and other parameters like $\alpha_{NN}$ and
$\beta_{NN}$ used to calculate the profile function.}
\begin{tabular}{cccccc}
\hline
\hline E(MeV/A) & $\sigma_{NN}$ $(fm^2)$& $\alpha_{NN}$ & $\beta_{NN}$$(fm^2)$\\
\hline
\hline
230.0 & 3.307790 & 0.7462136 & 0.1042526  \\
240.0 & 3.266868 & 0.6800303 & 0.0978437  \\
730.0 & 4.174130 & -0.0828693 & 0.1896611  \\
740.0 & 4.189708 & -0.0793203 & 0.1931483  \\
760.0 & 4.217336 & -0.0731153 & 0.1996571  \\
790.0 & 4.250772 & -0.0691061 & 0.2078210  \\
900.0 & 4.311690 & -0.1439280 & 0.2171934  \\
905.0 & 4.312775 & -0.1498595 & 0.2170294  \\
920.0 & 4.315433 & -0.1684008 & 0.2163436  \\
950.0 & 4.318554 & -0.2078482 & 0.2142974  \\
955.0 &4.318853  &-0.2146004  &0.2138945\\
960.0 & 4.319103 & -0.2213738 & 0.2134798  \\
965.0 & 4.319310 & -0.2281574 & 0.2130555  \\
980.0 & 4.319725 & -0.2484590 & 0.2117466 \\
1005.0 & 4.320055 & -0.2814692 & 0.2095773  \\
1010.0 & 4.320110 & -0.2878605 & 0.2091625  \\
1020.0 & 4.320220 & -0.3004108 & 0.2083566  \\
\hline \hline
\end{tabular}
\end{center}
\end{table}

\subsection{Total reaction cross section}
The main ingredient of the Glauber model is the densities of projectile and target nuclei.
The densities from the well known RMF with NL3 parameter are used from both axially deformed and
spherically symmetric formalisms along with the densities of
HF(SEI-I). As our earlier work increase curiosity to see the effect of simple effective
interaction on reaction dynamics.\
We need these densities in terms of the Gaussian coefficients
for the investigation of reaction dynamics. We converted these densities of HF(SEI), Sph. RMF(NL3) as well as
Def. RMF(NL3) in term of Gaussian coefficients as
\begin{equation}
\rho(r)=\sum_{i=1}^2 c_i exp[-a_i r^2].
\end{equation}
TABLE III presents
spherical and deformed densities from RMF(NL3) along with non relativistic HF(SEI-I) densities
in terms of their Gaussian coefficients $c_{i}$ and $a_{i}$.\\
Another important ingredient for evaluation of profile function in Glauber model
is energy dependent parameters $\sigma_{NN}$, $\alpha_{NN}$ and $\beta_{NN}$, where $\sigma_{NN}$
is the total nuclear reaction cross section in NN collision, $\alpha_{NN}$ is ratio of real to the imaginary
part of forward nucleon-nucleon scattering amplitude and $\beta_{NN}$ is slop parameter, which determines
the fall of the angular distribution of the NN scattering. These parameters are energy as well isospin dependent.
Table IV contains these energy dependent parameters over energy range (230-1020)MeV/A. These parameters
have been estimated by using the spline interpolation of values as suggested in Ref. \cite{horiu07}.

\begin{table*}
\renewcommand{\arraystretch}{0.9}
\tabcolsep 0.03cm
\caption{\label{tab:table5} Total nuclear reaction cross section with various
projectiles with $^{12}C$ target. The experimental data \cite{abu2000}
are given for comparison.}
\begin{tabular}{lccccccccc}
\hline
\hline
& Energy& \multicolumn{2}{r}{$\sigma_R(mb)$} &\multicolumn{5}{r}{$\chi_i^{2}$}\\
\cline{3-6}\cline{8-10}
Proj.&(A MeV)& HF(SEI-I) & RMF( NL3)& RMF(NL3) & Expt.& & HF(SEI-I)&RMF( NL3)&RMF( NL3)\\
&&(sph.)&(sph.)&(Def.) &&&(Sph.)&(Sph.)&(Def.)\\
\hline \hline
$^{9}Be$ & 790 & 857 &810 &844& 806$\pm$9&&3.227&0.020 &5.907 \\
$^{10}Be$ & 790 & 894 &836 &860& 813$\pm$10&&8.070 &0.651 &6.027 \\
$^{11}Be$ & 790 & 942 & 890&902& 942$\pm$8&&0.000 &2.870 &0.000\\
$^{12}Be$ & 790 & 986 &894 &1004& 927$\pm$18&&3.755&1.175&1.247 \\
$^{12}B$ & 790 & 972&875 &909& 866$\pm$7&&12.975&0.094 &3.621 \\
$^{13}B$ & 790 & 1012 &916 &960& 883$\pm$14&&18.846&1.233 &7.248\\
$^{14}B$ & 790 & 1049 & 951&1029& 929$\pm$26&&15.501&0.521 &10.338\\
$^{15}B$ & 790 & 1086 &991 &1080& 962$\pm$160&&15.983&0.874&21.856\\
$^{13}C$ & 960 & 1005&901 &931 & 862$\pm$12&&23.723&1.765&5.208\\
$^{14}C$ & 965 & 1003&944 & 978& 880$\pm$19&&17.192&4.655&10.473\\
$^{15}C$ & 730 & 1065 &959 &1026& 945$\pm$10&&15.238&0.207&5.795\\
$^{16}C$ & 960 & 1113&1005&1084& 1036$\pm$11&&5.723&0.928&3.831\\
$^{17}C$ & 965 & 1147&1037&1117& 1056$\pm$10&&7.842&0.342&5.910\\
$^{18}C$ & 955 & 1182&1098 &1150 & 1104$\pm$15&&5.511&0.033&4.066\\
$^{19}C$ & 960 & 1215 &1099 &1120& 1231$\pm$28&&0.208&14.154&0.007\\
$^{20}C$ & 905 & 1271 &1131 &1179& 1187$\pm$20&&5.944&2.642&3.033\\
$^{20}N$ & 950 & 1239 &1141 &1174& 1121$\pm$17&&12.421&0.357&1.031\\
$^{21}N$ & 1005 & 1270 &1172 &1195& 1114$\pm$9&&21.846&3.020&13.581\\
$^{22}N$ & 965& 1306 &1205 & 1278& 1245$\pm$49&&2.989&1.285&0.001\\
$^{23}N$ & 920 & 1344 &1240 & 1483&1399$\pm$98&&2.162&18.071&8.807\\
$^{20}O$ & 950 & 1233 &1130 &1179& 1078$\pm$10&&22.287&2.508&7.348 \\
$^{21}O$ & 980 & 1261 &1158 &1197& 1098$\pm$11&&24.198&3.279&8.219 \\
$^{22}O$ & 965 & 1291 &1188 &1225 & 1172$\pm$22&&12.083&0.218&2.049 \\
$^{23}O$ & 960 & 1327 &1218 &1257 & 1308$\pm$16&&0.276&6.193&2.398 \\
$^{24}O$ & 965 & 1367 &1251 &1292 & 1318$\pm$52&&1.822&3.406&0.777 \\
$^{23}F$ & 1020 & 1316 &1211 &1250 & 1148$\pm$16&&24.585&3.457&8.537 \\
$^{24}F$ & 1005& 1354 &1242 &1299& 1253$\pm$23&&8.141&0.097&0.352 \\
$^{25}F$ & 1010& 1393 &1276 &1363 & 1298$\pm$31&&6.953&0.373&0.093 \\
$^{26}F$ & 950 & 1434 &1315 & 1388& 1353$\pm$54&&4.849&1.067&0.018 \\
$^{28}Ne$ & 240 & 1382&1272 &1283& 1273$\pm$11&&9.333&0.001&4.186 \\
$^{29}Ne$ & 240 & 1417&1302 &1340& 1344$\pm$14&&3.965&1.313&0.003 \\
$^{28}Ne$ & 950 & 1497&1378 &1391& 1244$\pm$40&&51.454&14.434&36.814 \\
$^{29}Ne$ & 950 & 1534&1410 &1452& 1279$\pm$32&&50.841&13.418&23.400 \\
$^{30}Ne$ & 240 & 1454&1339 &1401& 1348$\pm$17&&8.335&0.060&0.012 \\
$^{31}Ne$ & 240 & 1481&1353 &1408& 1435$\pm$22&&1.475&4.686&0.001 \\
$^{32}Ne$ & 240 & 1509&1375 &1417& 1385$\pm$33&&11.102&0.072&16.463 \\
$^{32}Mg$ & 900 & 1613&1482 &1545& 1331$\pm$24&&59.748&17.131&25.437 \\
$^{32}Mg$ & 950 & 1613& 1482&1545& 1340$\pm$24&&55.619&15.048&76.418 \\
$^{33}Mg$ & 900 & 1641&1505 &1538& 1320$\pm$23&&78.061&25.928&37.001 \\
$^{34}Mg$ & 900 & 1669&1526 &1534& 1372$\pm$46&&64.292&17.286&59.202 \\
$^{35}Mg$ & 900 & 1697&1548 &1553& 1472$\pm$70&&34.392&3.924&37.198 \\
\hline
\hline
\end{tabular}
\end{table*}
The calculated values of reaction cross sections for considered nuclei using the
HF(SEI), Sph. RMF(NL3)and Def. RMF(NL3) densities are presented in the TABLE V.
The calculated vales of reaction cross sections are also compared with the experimental
data. It seen from the TABLE, the $\sigma_R$ obtained from both relativistic and non-relativistic
formalisms are well comparable with the experimental observation. To analysis of the observation
we used the $\chi^2$ fitting. The chi-square fitting ($\chi^2$-fitting) can be defined as
\begin{equation}
\chi^2 = \sum\chi_i^2 = \sum_i \frac{(\sigma_R(Obs)-\sigma_R(Exp))^2}{\sigma_R(Exp)}.
\end{equation}
The table also presents the $\chi_i^2$ fitting values of $\sigma_R$ for various densities with experimental observations.
The overall $\chi^2$ fitting values for HF(SEI), spherical RMF(NL3) and
Deformed RMF(NL3) densities are 753.157, 189.335 and 493.407 respectively. The values of $\sigma_R$ for
the $^{11}$Be, $^{15}$C, $^{19}$C, $^{23}$O and $^{31}$Ne halo nuclei are 942, 1065, 1215, 1327, 1481 in mb
for HF(SEI), 902, 1026, 1120, 1257, 1408 in mb for Def. RMF(NL3), 890, 959, 1099, 1218,
1353 in mb for Sph. RMF(NL3) densities and 942$\pm$8, 945$\pm$10, 1231$\pm$28, 1308$\pm$16, 1435$\pm$22 in mb from the experimental data.
The value of $\chi^2$ is small for the Sph. RMF(NL3) density shows, the
$\sigma_R$ obtained with this density reproduced better results as compare to the other cases.

\begin{figure*}
\includegraphics[width=2.2\columnwidth,clip=true]{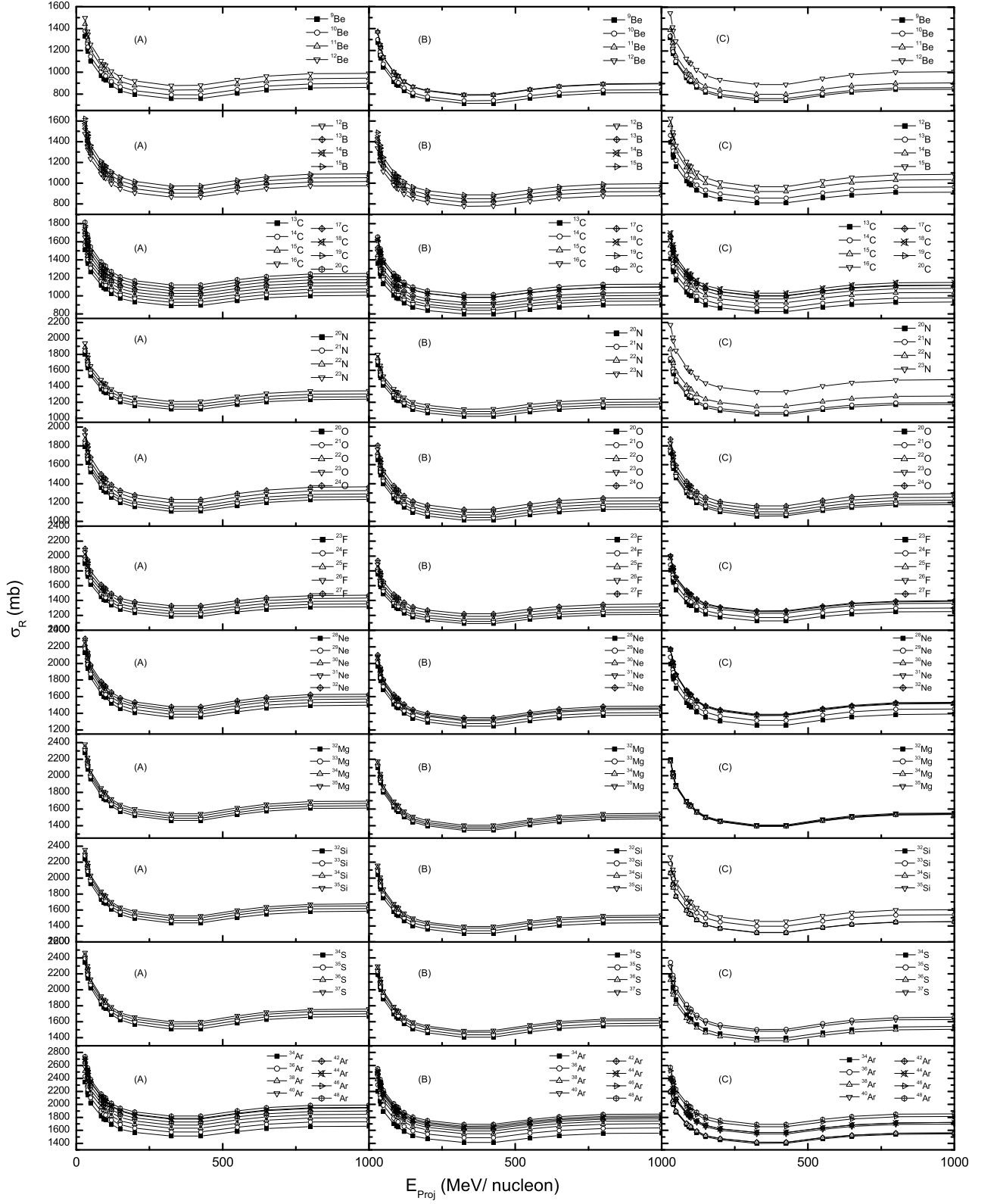}
\caption{Variation of total reaction cross sections ($\sigma_R$) as a projectile energy ($E_{proj}$) using (A) HF(SEI),
(B) Sph. RMF(NL3), (C) Def. RMF(NL3) densities for isotopes of $^{9-12}$Be, $^{12-15}$B, $^{13-20}$C, $^{20-23}$N, $^{20-24}$O,
$^{23-27}$F, $^{28-32}$Ne, $^{32-35}$Mg, $^{32-35}$Si, $^{34-37}$S and $^{34-48}$Ar nuclei.} \label{Fig.1}
\end{figure*}

 By looking
deep into the table one can easily interpret that the values of reaction cross sections obtained for particular
halo systems and other exotic nuclei shows better results with deformed RMF(NL3) and HF(SEI) densities. \\
Fig. 4 shows the total reaction cross sections of $^{9-12}$Be, $^{12-15}$B, $^{13-20}$C, $^{20-23}$N, $^{20-24}$O,
$^{23-27}$F, $^{28-32}$Ne, $^{32-35}$Mg, $^{32-35}$Si, $^{34-37}$S and $^{34-48}$Ar nuclei as a function of projectile energy ($E_{proj}$) over
the energy range of (30-1000) MeV/Nucleon. The value of $\sigma_R$ has higher value at small $E_{proj}$ and start decreasing
upto the energy range of 300 MeV/nucleon. Small enhancement in $\sigma_R$ is observed upto the range of 750 MeV/nucleon and
after that, it remains constant.
The observation from the table shows the values of $\sigma_R$ obtained from the HF(SEI) densities are
slightly higher than the other densities. The value of $\sigma_R$ also increases by equal proportion in their isotopic chain
as a mass number for both cases of HF(SEI) and Sph. RMF(NL3) densities. But some disorder in increase of the $\sigma_R$ is
appeared for the case Def. RMF(NL3) densities. This effect may be because of the role of deformation in $\sigma_R$. \\

\section {Summary and Conclusions}
In summary, we study the structural properties and $\sigma_R$ for $^{9-12}$Be, $^{12-15}$B, $^{13-20}$C, $^{20-23}$N, $^{20-24}$O,
$^{23-27}$F, $^{28-32}$Ne, $^{32-35}$Mg, $^{32-35}$Si, $^{34-37}$S and $^{34-48}$Ar nuclei using the densities from the non relativistic
Hartree-Fock with simple effective interaction as well as relativistic mean field formalism. The bulk properties like binding
energy, charge radius and quadrupole deformation parameter $\beta_2$ are studied with these formalisms which shows good agreement with the
experimental data. Bubble effects for some of the nuclei like $^{22}$O, $^{23}$F, $^{34}$Si and $^{46}$Ar are appeared by our study.
Such effects are required on nuclear structure and reaction studies in dripline as well as superheavy region.
The $\sigma_R$ obtained with the RMF formalism shows better results as compare to the non relativistic
HF(SEI) densities for general cases. But the remarkable agreement is appeared with HF(SEI) densities for the halo nuclear
systems as compare to RMF densities. In general, we found both the formalisms are equally capable for the study of structure and
reaction dynamics for most of the nuclear systems in the light mass region. It needs further investigations to see effect of these
formalisms on other nuclear systems in both medium and heavy mass regions.

\begin{acknowledgments}
One of the author Mahesh K. Sharma thanks the institute of Physics, Bhubaneswar for their kind hospitality.

\end{acknowledgments}

\end{document}